\documentclass[journal,draftclsnofoot,onecolumn,12pt]{IEEEtran}

\newif\ifarxiv
\arxivtrue 

\newif\ifonecolumn
\onecolumntrue 

\IEEEoverridecommandlockouts
\usepackage{multirow}
\usepackage{dsfont}
\usepackage{cite}
\usepackage{amsmath,amssymb,amsfonts, mathtools}
\usepackage{float}
\usepackage{makecell}
\usepackage{caption}
\usepackage{graphicx}
\usepackage{textcomp}
\usepackage{algorithm}
\usepackage[left=0.625in,right=0.625in,bottom=1.0in,top=0.75in,bindingoffset=0mm]{geometry}
\usepackage{amsthm}
\usepackage{algpseudocode}
\usepackage{adjustbox}
\usepackage{xcolor}
\usepackage{verbatim}
\usepackage{paralist}
\usepackage{array}
\newcolumntype{?}{!{\vrule width 1.1pt}}
\usepackage{tablefootnote}

\theoremstyle{remark}

\definecolor{green}{rgb}{0.0, 0.5, 0.0} 
\allowdisplaybreaks
\usepackage{acro}
\newcolumntype{?}{!{\vrule width 1pt}}

\DeclareAcronym{snr}{
  short = SNR,
  long = signal-to-noise ratio,}
  \DeclareAcronym{pdf}{
	short = PDF,
	long = probability density function,}
  \DeclareAcronym{sar}{
  short = SAR,
  long = synthetic aperture radar,}

\DeclareAcronym{insar}{
  short = InSAR,
  long = interferometric synthetic aperture radar,}

\DeclareAcronym{ao}{
	short = {AO},
	long = {alternating optimization},
	long-plural-form = {alternating optimizations}
}
\DeclareAcronym{mimo}{
	short = {MIMO},
	long = {multiple-input multiple-output}
}

\DeclareAcronym{uav}{
        short = {UAV},
        long = {unmanned aerial vehicle},
        long-plural-form = {unmanned aerial vehicles}
}

\DeclareAcronym{fdma}{
	short = {FDMA},
	long = {frequency-division multiple-access},
}
\DeclareAcronym{1d}{
	short = {1D},
	long = {one-dimensional},
}
\DeclareAcronym{islr}{
	short = {ISLR},
	long = {integrated sidelobe ratio},
}

\DeclareAcronym{pslr}{
	short = {PSLR},
	long = {peak sidelobe ratio},
}

\DeclareAcronym{3d}{
        short = {3D},
        long = {three-dimensional},
}
\DeclareAcronym{pso}{
	short = {PSO},
	long = {particle swarm optimization},
}

\DeclareAcronym{2d}{
        short = {2D},
        long = {two-dimensional},
}

\DeclareAcronym{dem}{
        short = {DEM},
        long = {digital elevation model},
}
\DeclareAcronym{gs}{
        short = {GS},
        long = {ground station},
        long-plural-form = {ground stations}
}

\DeclareAcronym{los}{
        short = {LOS},
        long = {line-of-sight},
}

\DeclareAcronym{sca}{
        short = {SCA},
        long = {successive convex approximation},
}

\DeclareAcronym{nesz}{
        short = {NESZ},
        long = {noise equivalent sigma zero},
}

\DeclareAcronym{wrt}{
        short = {w.r.t.},
        long = {with respect to},
}
\DeclareAcronym{rhs}{
        short = {r.h.s},
        long = {right-hand side},
}
\DeclareAcronym{gmti}{
	short = {GMTI},
	long = {ground moving target indication},
}
\DeclareAcronym{lhs}{
        short = {l.h.s},
        long = {left-hand side },
}
\DeclareAcronym{bcd}{
	short = {BCD},
	long = {block coordinate descent},
}
\DeclareAcronym{hoa}{
        short = {HoA},
        long = {height of ambiguity},
}
\DeclareAcronym{stso}{
	short = {STSO},
	long = {short-term shift-orthogonal},
}
\DeclareAcronym{tomosar}{
	short = {TomoSAR},
	long = {tomographic SAR},
}
\DeclareAcronym{tdbp}{
	short = {TDBP},
	long = {time-domain back projection},
}
\DeclareAcronym{iid}{
	short = {i.i.d.},
	long = {independent and identically distributed},
}
\DeclareAcronym{awgn}{
	short = {AWGN},
	long = {additive white gaussian noise},
}
\DeclareAcronym{psf}{
	short = {PSF},
	long = {point spread function},
}
\DeclareAcronym{aoi}{
	short = {AoI},
	long = {area of interest},
}
\DeclareAcronym{ea}{
	short = {EA},
	long = {evolutionary algorithm},
}
\DeclareAcronym{ga}{
	short = {GA},
	long = {genetic algorithm},
	long-plural-form = {genetic algorithms},
}

\DeclareAcronym{cga}{
	short = {CGA},
	long = {Continuous Genetic Algorithm},
	long-plural-form = {continuous genetic algorithms},
}

\DeclareAcronym{genocop}{
	short = {Genocop II},
	long = {Genetic Algorithm for Numerical Optimization of Constrained Problems},
}

\DeclareAcronym{psl}{
	short = {PSL},
	long = {peak sidelobe level},
}
\DeclareAcronym{ula}{
	short = {ULA},
	long = {uniform linear array},
}

\begin{document}
\title{Point Spread Function Optimization for Communication-Assisted UAV-Borne MIMO TomoSAR \\\vspace{-2mm}
\thanks{  This work was supported in part by the Deutsche Forschungsgemeinschaft (DFG, German Research Foundation) GRK 2680 – Project-ID 437847244.}
}
\author{\IEEEauthorblockN{Pouya Fakharizadeh\IEEEauthorrefmark{1}, Mohamed-Amine~Lahmeri\IEEEauthorrefmark{1}, Gerhard Krieger\IEEEauthorrefmark{1}\IEEEauthorrefmark{2}, and
Robert Schober\IEEEauthorrefmark{1}}\\ \vspace{-2mm}
\IEEEauthorblockA{\IEEEauthorrefmark{1}Friedrich-Alexander-Universit\"at Erlangen-N\"urnberg (FAU), Germany\\
\IEEEauthorrefmark{2}German Aerospace Center (DLR),  Microwaves and Radar Institute, Weßling, Germany\\
\vspace{-8mm}}}
\maketitle
\begin{abstract} 
  This paper tackles the optimization of {the} \acf{psf} of \acf{uav}-borne \acf{mimo} \acf{sar} tomography systems. A swarm of UAV-borne \ac{sar} systems is deployed to image an area to obtain its height profile. To achieve a high-quality \acf{3d} image of the scene, the \ac{psf} has to exhibit low sidelobes. The heavy computations {required for image generation} are performed on the ground. {To this end}, the sensor data collected by the \ac{uav}-\acp{sar} is offloaded in real time via a frequency division multiple access (FDMA) air-to-ground backhaul link. In this work, the \ac{uav} formation and the power allocated for offloading are jointly optimized for the minimization of the \ac{psf} sidelobe levels. To this end, we propose a novel solution based on \acf{pso}, which meets practical sensing and communication constraints. Our simulation results demonstrate that the proposed solution can {significantly improve sidelobe suppression compared to several benchmark schemes.}
\end{abstract}
\section{Introduction}
The deployment of \acf{uav} swarms for remote sensing has garnered considerable attention in recent years. This growing interest has facilitated the broader application of \acp{uav} across various domains, including mapping, traffic monitoring, and climate change analysis. To support these applications, \ac{uav} platforms are equipped with a range of sensors, including optical cameras, LiDAR systems, and radar instruments. Notably, the integration of \acf{sar} into \ac{uav} systems has emerged as a particularly promising option, as it enables the generation of high-resolution imagery of local areas, even under adverse environmental conditions, thereby overcoming the constraints associated with conventional airborne and spaceborne \ac{sar} platforms. 

An important remote sensing application {of} \ac{uav} swarms is \acf{3d} radar imaging. 
In particular, \ac{tomosar} exploits a synthetic aperture in elevation to retrieve the vertical distribution of scatterers \cite{tutorial}. One key performance metric in \ac{tomosar} is the \acf{psf}, which characterizes how a scatterer in elevation is mapped onto the reconstructed tomographic image \cite{fornaro}. Although the \ac{psf} is well established in the radar literature for Fourier-based tomography, where typically far-field conditions and uniform platform spacing are assumed \cite{reigber}, its extension to \ac{uav}-based systems with arbitrary platform formation has remained largely unexplored. The \ac{psf} determines essential imaging properties, including tomographic resolution, height of ambiguity, and sidelobe levels, all of which directly influence the quality of the resulting image. In particular, the sidelobe levels quantify the extent to which energy from a strong scatterer leaks into adjacent elevation bins. Strong sidelobes can mask weaker scatterers, generate ghost targets, and ultimately degrade the interpretability of complex scenes, such as urban or volumetric environments \cite{zhu}. Other relevant metrics for \ac{tomosar} include the \ac{snr} and coverage \cite{kreiger}. While \ac{tomosar} has been extensively studied for spaceborne and airborne platforms \cite{reigber}, the optimization of \ac{tomosar} performance for \ac{uav}-borne systems has remained unexplored. Furthermore, \acf{mimo} \ac{tomosar} offers several advantages over conventional repeat-pass \ac{tomosar} \cite{reigber}, in which a single \ac{sar} platform performs multiple acquisitions of the same scene. Notably, it enables improved discrimination between single-bounce and double-bounce scattering mechanisms \cite{kreiger} and exhibits enhanced robustness to pronounced sidelobes \cite{seker}. The optimization of \ac{uav}-based \ac{insar} was considered in \cite{amine3,amine5,amine6}.  However, these results are not directly applicable here, as \ac{insar} relies on acquisition pairs. Consequently, it cannot recover the scatterer distribution along the elevation dimension and is therefore unable to resolve multiple scatterers within the same resolution cell. In contrast, \ac{tomosar} is capable of reconstructing the full \ac{3d} image by using multiple acquisitions, thereby enabling the separation of scatterers located at different heights but projected onto the same resolution cell \cite{rambour}.

In this work, we study a \ac{uav}-borne \ac{mimo} \ac{tomosar} system, where a swarm of \acp{uav}, which transmit and receive radar waveforms, offloads the collected radar data to a \ac{gs} to generate a \ac{3d} image of an area. We investigate the joint optimization of the \ac{uav} formation and communication power allocation for minimization of the \acf{psl} of the \ac{psf} under communication and sensing constraints. Our main contributions can be summarized as follows:\begin{itemize}
\item We jointly optimize the \ac{uav} formation and communication power allocation for minimization of the \ac{psl} of the \ac{psf}, while satisfying sensing and communication constraints.
\item We propose a \acf{pso} algorithm to solve the formulated optimization problem.
\item Our simulation results demonstrate the effectiveness of the proposed approach compared to  several benchmark schemes, achieving \acp{psl} between $-17$ dB and $-33$ dB.
\end{itemize}
{\em Notations}:
In this paper, lower-case letters $x$ refer to scalar {variables}, while boldface lower-case letters $\mathbf{x}$ denote vectors.  $\{a, ..., b\}$ denotes the set of all integers between $a$ and $b$.  $|x|$ denotes the absolute value of $x$. $\mathbb{R}^{N}$ represents the set of all $N$-dimensional vectors with real-valued entries. For a vector {$\mathbf{x}=(x_1,...,x_N)^{\rm T}\in\mathbb{R}^{N}$}, $||\mathbf{x}||_2$ denotes the Euclidean norm, whereas  $\mathbf{x}^{\rm T }$ stands for the  transpose of $\mathbf{x}$. $[x]^+, x\in\mathbb{R}$, denotes $\max(x,0)$. $\mathcal{U}(\mathbf{a},\mathbf{b})$ refers to a uniform distribution between the elements of $\mathbf{a}$ and $\mathbf{b}$. $\mathbf{x}\odot\mathbf{y}$ denotes the element-wise multiplication of the elements of vectors $\mathbf{x}$ and $\mathbf{y}$.
\section{System Model} \label{Sec:SystemModel}
\begin{figure}
	\centering
	\ifonecolumn
	\includegraphics[width=4in]{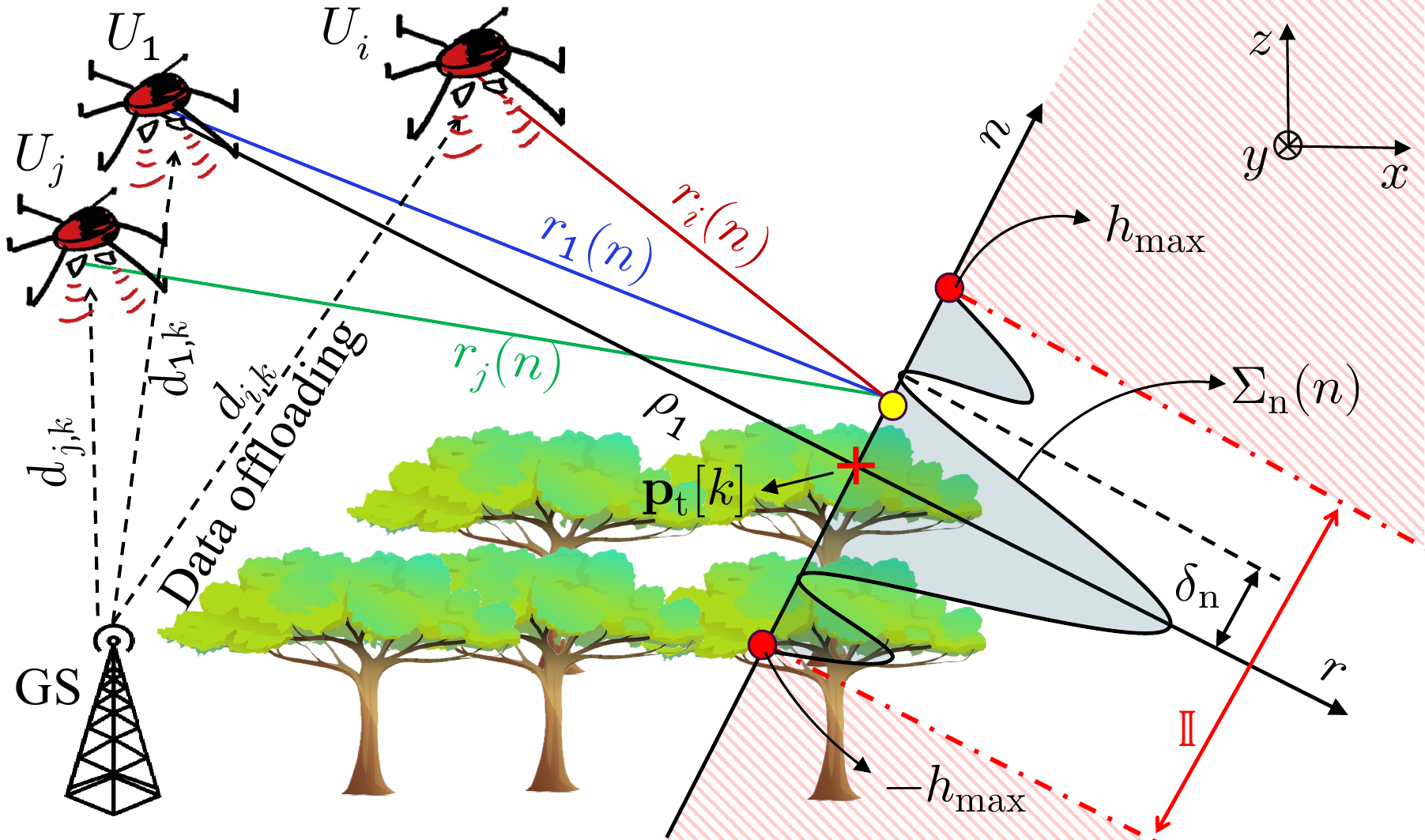}
	\else
	\includegraphics[width=0.9\columnwidth]{figures/MIMO_SAR_config_2}
	\fi
	\caption{ \ac{uav}-borne \ac{mimo} \acs{tomosar} sensing system comprising $I$ \ac{uav}-\ac{sar} sensors and the \ac{gs} for real-time data offloading. The \ac{uav} swarm's flight direction is along the $y$-axis. The elevation \ac{psf} $\Sigma_{\rm n}(n)$ is evaluated on the $n$-axis over the interval $\mathbb{I}$.}
	\label{fig:system-model}
\end{figure} 
We consider a set of $I\in \mathbb{N}$, $I\geq2$, rotary-wing \acp{uav}, {denoted} by $U_i, i \in\mathcal{I}=\{1,...,I\}$, performing \ac{tomosar} over an area, see Fig.~\ref{fig:system-model}. Each \ac{uav} transmits radar waveforms and receives not only the echoes from its own transmission but also those originating from other \acp{uav}, thereby forming a \ac{mimo} system. A \ac{3d} Cartesian coordinate system is adopted, where the  $x$-,  $y$-, and $z$-{axes} correspond to the range, azimuth, and altitude directions, respectively. The total mission time $T$ is discretized into $K$ slots {of equal duration $\delta_{\rm t}$}, such that $T = K\cdot\delta_{\rm t}$, and the set of time indices is defined as  $\mathcal{K}=\{1,...,K\}$. The considered \ac{uav}-\ac{sar} system operates in stripmap mode \cite{book1}, with all \acp{uav} maintaining a constant velocity $v_{\rm y}$ along linear trajectories parallel to reference line $l_{\rm t}$, {which} is aligned with {the} $y$-axis. In time slot $k$, this reference line passes through the point $\mathbf{p}_{\rm t}[k]=(x_{\rm t},y[k],0)^{\rm T} \in \mathbb{R}^3$, see Fig.~\ref{fig:system-model}. The position of $U_i$ {in} time slot $k \in \mathcal{K}$ is given by $\mathbf{q}_i[k]=(x_i,y[k],z_i)^{\rm T}$, where the azimuthal position vector $\mathbf{y}=(y[1]=0,y[2], ..., y[K])^{\rm T}\in\mathbb{R}^{K}$ is defined as $y[k+1]=y[k]+v_{\rm y}\delta_{\rm t}, \forall k \in \mathcal{K}\setminus\{K\}$.
For simplicity, we denote the position of $U_i$ in the across-track plane (i.e., the $xz-$plane) by $\mathbf{q}_i=(x_i,z_i)^{\rm T} \in \mathbb{R}^2, \forall i \in \mathcal{I}$. The distance between \acp{uav} $U_i$ and $U_j$ is given by $b_{i,j} = \left\|\mathbf{q}_i -\mathbf{q}_j\right\|_2, \forall i,j \in \mathcal{I}, i>j$. Without loss of generality, $U_1$ is selected as the reference \ac{uav} to define a local coordinate system, where the $r$- and $n$-axes correspond to the slant-range and normal directions, respectively, see Fig.~\ref{fig:system-model}. The $r$-axis is defined as  the line connecting $U_1$ to $\mathbf{p}_{\rm t}[k]$, while the $n$-axis lies in the $xz$-plane and is orthogonal to the $r$-axis and passes through $\mathbf{p}_{\rm t}[k]$.

 \subsection{\acs{tomosar} Performance Metrics}
Next, we introduce the relevant \ac{tomosar} performance metrics.
\subsubsection{\ac{sar} Coverage}
\ifonecolumn
Let $\rho_i, i \in \mathcal{I}$, {denote} $U_i$'s slant range \ac{wrt} $\mathbf{p}_{\rm t}[k]$. The slant range is independent of time and is given by:
\begin{align*}
	\rho_i= \sqrt{ (x_i -x_{\rm t})^2 +  z_i^2  }, \forall i \in \mathcal{I}.
\end{align*}

The radar swath is designed to be centered \ac{wrt} $l_{\rm t}$. To this end, the look angles of the \acp{uav}, denoted by $\theta_i, i \in \mathcal{I}$, are adjusted {such that} the beam footprints are centered at $\mathbf{p}_{\rm t}[k], \forall k$, i.e., $\theta_i= \arctan\left(\frac{x_{\rm t}-x_i}{z_i}\right)$. The swath width of $U_i$ can be approximated as follows \cite{book1}:
\begin{align*}\label{eq:swath_width}
	S_i = \frac{\Theta_{\rm 3 dB}\rho_i }{\cos\theta_i},\forall i \in \mathcal{I},
\end{align*}
where $\Theta_{\mathrm{3dB}}$ is the -3 dB beamwidth in elevation.
\else
The radar swath is designed to be centered \ac{wrt} $l_{\rm t}$. To this end, the look angles of the \acp{uav}, denoted by $\theta_i, i \in \mathcal{I}$, are adjusted {such that} the beam footprints are centered at $\mathbf{p}_{\rm t}[k], \forall k$, i.e., $\theta_i= \arctan\left(\frac{x_{\rm t}-x_i}{z_i}\right)$. 

The swath width of $U_i$ can be approximated as follows \cite{book1}:
\begin{align*}\label{eq:swath_width}
	S_i = \frac{\Theta_{\rm 3 dB}\rho_i }{\cos\theta_i},\forall i \in \mathcal{I},
\end{align*}
where $\Theta_{\mathrm{3dB}}$ is the -3 dB beamwidth in elevation and $\rho_i= \sqrt{ (x_i -x_{\rm t})^2 +  z_i^2  }, \forall i \in \mathcal{I}$, denotes $U_i$'s slant range \ac{wrt} $\mathbf{p}_{\rm t}[k]$.\fi
{\subsubsection{Point Spread Function (PSF)}{\label{subsection:mimo psf}}
\ac{tomosar}'s capability to suppress undesired scattering contributions and avoid distortions of the desired target is captured by the \ac{psf}\ifarxiv. We derive an expression for the \ac{mimo} \ac{tomosar} elevation \ac{psf} along the $n$-axis assuming small bi-static angles $|\theta_i-\theta_j|, \forall i, j$, on the order of a few degrees, which holds for \ac{tomosar} applications \cite{seker}. Based on this assumption, the radar cross section is independent of $\theta_i,\forall i$.
The noise-free radar signal received at $U_i$ from all \acp{uav} $U_l, l\in\mathcal{I}$, reflected by a point target on the $n$-axis with local coordinates $(0,n)^{\rm T}$ is given by \cite{fornaro2}:
\begin{equation*}
	y_i = \sigma\sum_{l=1}^{I}\alpha_{il}\exp\left\{j\frac{2\pi}{\lambda}(r_i(n)+r_l(n))\right\},\forall i \in \mathcal{I},
\end{equation*}
where $\sigma$ represents the backscattering coefficient along the $n$-axis and $\alpha_{il}>0, \forall i,l\in\mathcal{I}$, is the attenuation coefficient associated with the path loss from $U_l$ to $U_i$. $\lambda$ is the wavelength of the radar signal and $r_i(n)$ denotes the slant range from $U_i$ to the respective point located on the $n$-axis (see Fig.~\ref{fig:system-model}) and is given by:
\begin{equation*}
	r_i(n)=\sqrt{\left(x_i-\left(x_{\rm t}+n\cos\theta_1\right)\right)^2+\left(z_i-n\sin\theta_1\right)^2},\forall i \in \mathcal{I}.
\end{equation*}
 We can write the sum of all received signals normalized by the number of acquisitions, assuming $\alpha_{il}=\alpha,\forall i,l\in\mathcal{I}$, (i.e., after proper normalization of the transmit and received signals) as follows:
\begin{align*}
	y&=\frac{1}{I^2}\sum_{i=1}^{I}y_i=\frac{\alpha\cdot\sigma}{I^2}\sum_{i=1}^{I}\sum_{l=1}^{I}\exp\left\{j\frac{2\pi}{\lambda}(r_i(n)+r_l(n))\right\}\\
	&=\alpha\cdot\sigma\cdot\underbrace{\frac{1}{I^2}\left(\sum_{i=1}^{I}\exp\left\{j\frac{2\pi}{\lambda}r_i(n)\right\}\right)^2}_{\text{\ac{tomosar} response}},\forall i \in \mathcal{I}.
\end{align*}
The \ac{psf}, denoted by $\Sigma_{\rm n}(n)$, is defined as the magnitude of the \ac{tomosar} response \cite{fornaro2} and is given by:  
\begin{equation}
	\Sigma_{\rm n}(n)=\frac{1}{I^2}\left|\sum_{i=1}^{I}\exp\left\{j\frac{2\pi}{\lambda}r_i(n)\right\}\right|^2,\forall i \in \mathcal{I}.
	\label{eq:psf}
\end{equation}
Using a similar procedure, we can derive the range \ac{psf} along the $r$-axis as follows:
  \begin{equation}
  	\Sigma_{\rm r}(r)=\frac{1}{I^2}\left|\sum_{i=1}^{I}\exp\left\{j\frac{2\pi}{\lambda}l_i(r)\right\}\right|^2,\forall i \in \mathcal{I},
  	\label{eq:psf_r}
  \end{equation}
  where $l_i(r)=\sqrt{\left(x_i-\left(x_{\rm t}+r\sin\theta_1\right)\right)^2+\left(z_i+r\cos\theta_1\right)^2},\forall i \in \mathcal{I}$, 
  denotes the distance between $U_i$ and a point located on the $r$-axis with local coordinates $(r,0)^{\mathrm{T}}$. Since the primary objective of \ac{tomosar} is the reconstruction of the height profile, in the remainder of this work, we focus on the function $\Sigma_{\rm n}(n)$ to characterize the height resolution and the \ac{psl}. $\Sigma_{\rm r}(r)$ is employed to evaluate the slant-range resolution.
\fi}
\subsubsection{\ac{tomosar} Resolution}
Key performance metrics for \ac{tomosar} are the tomographic and slant-range resolutions, denoted by $\delta_{\rm n}$ and $\delta_{\rm r}$, respectively. These metrics characterize the system's ability to distinguish scatterers along the $n$- and $r$-axes. In this work, we adopt the peak-to-null Rayleigh resolution \cite{seker}. Accordingly, the tomographic and slant-range resolutions are derived from their respective \acp{psf} as follows:
\ifonecolumn
\begin{equation} \label{eq:resolutions}
    \delta_{\Delta}= \frac{1}{2}\left(\underbrace{\min\limits_{ \Delta > 0 } \left\{\; \Delta \middle| \; \Sigma_{\rm \Delta}(\Delta)=0 
\right\}}_{\text{first positive null}}- \underbrace{\max\limits_{ \Delta< 0 } \left\{\; \Delta \middle| \; \Sigma_{\rm \Delta}(\Delta)=0
\right\}}_{\text{first negative null}}\right), \forall \Delta \in \{\rm n,\rm r\}.
\end{equation}\else
{\small\begin{gather}
 \label{eq:resolutions}
    \delta_{\Delta}= \frac{1}{2}(\underbrace{\min\limits_{ \Delta > 0 } \left\{\; \Delta \middle| \; \Sigma_{\rm \Delta}(\Delta)=0 
\right\}}_{\text{first positive null}}- \underbrace{\max\limits_{ \Delta< 0 } \left\{\; \Delta \middle| \; \Sigma_{\rm \Delta}(\Delta)=0
\right\}}_{\text{first negative null}}),\\ \notag
\forall \Delta \in \{\rm n,\rm r\}.
\end{gather}}\fi
\subsubsection{Peak Sidelobe Level (PSL)}
Since the tomographic height profile is characterized by the \ac{psf} along the $n$-axis, we focus on optimizing $\Sigma_{\rm n}(n)$. To quantify the sidelobes relevant for the height reconstruction, we define a symmetric interval, $\mathbb{I}=[-h_{\mathrm{max}}, h_{\mathrm{max}}]$, along the $n$-axis. Within this interval, the main lobe is defined to reside in $\left[-{\delta_{\rm n}}, {\delta_{\rm n}}\right]$, whereas the remaining region corresponds to sidelobes, see Fig.~\ref{fig:system-model}. Accordingly, the \ac{psl} is defined as:
\begin{equation}
	\mathrm{PSL}=\max_{n\in\mathbb{I}\setminus[-{\delta_{\rm n}}, {\delta_{\rm n}}]}\Sigma_{\rm n}(n).
    \label{eq:psl}
\end{equation} 
Note that $h_{\mathrm{max}} \in \mathbb{R}$ is practically designed to meet the \ac{tomosar} requirement on the minimum height of ambiguity, which is determined by the location of the first replica of $\Sigma_{\rm n}(n)$ \cite{seker}. Therefore, the choice of $h_{\mathrm{max}}$ affects the maximum height that can be reliably reconstructed without ambiguity. 
\subsubsection{Signal-to-Noise Ratio (\ac{snr})}
The \ac{snr} is another determining factor for the quality of the final \ac{sar} image. The received \ac{snr} from $U_i$ to $U_j$ is given by \cite{book1}:
\begin{equation*}
	\mathrm{SNR}_{i,j}=\alpha \left(\frac{1}{\rho_i\rho_j}\right)^2, \forall i,j\in\mathcal{I},
\end{equation*}
where $\alpha=\frac{\sigma_0 A_{\mathrm{res}} P_{\rm t} G_{\rm t} G_{\rm r} \lambda^2}{(4 \pi)^3 k_b T_{\mathrm{sys}} B_{\mathrm{noise}} F L}$. Here, $\sigma_0$ denotes the normalized backscattering coefficient and $A_{\mathrm{res}}= \delta_{\rm r}\cdot \delta_{\rm n}$ represents the resolution cell size, i.e., the smallest area that the radar can distinguish. $P_{\rm t}$ is the radar transmit power for all $U_i$s, $G_{\rm t}$ and $G_{\rm r}$ are the transmit and receive antenna gains for all $U_i$s, respectively, $k_b$ is the Boltzmann constant, $T_{\mathrm{sys}}$ is the receiver temperature, $B_{\mathrm{noise}}$ is the effective noise bandwidth, $F$ is the noise figure, and  $L$ represents the total radar losses.

\subsection{Communication Performance}
We consider real-time offloading of the radar data to the \ac{gs}, see Fig.~\ref{fig:system-model}, where the \acp{uav} employ \ac{fdma}. The communication transmit power consumed by $U_i$ is given by $\mathbf{P}_i=(P_{i,1},...,P_{i,K})^{\rm T} \in \mathbb{R}^K, i \in \mathcal{I}$.
We denote the location of the \ac{gs} by $\mathbf{g}= (g_{\rm x}, g_{\rm y}, g_{\rm z})^{\rm T} \in \mathbb{R}^3$ and the distance from $U_i$ to the \ac{gs} by    $d_{i,k}(\mathbf{q}_i) = ||\mathbf{q}_i[k]-\mathbf{g} ||_2, \forall i \in \mathcal{I}, \forall k \in \mathcal{K}.$ Thus, {adopting} the free-space path loss model and \ac{fdma}, the
instantaneous throughput  from $U_i$ to the \ac{gs} is given by:
\begin{align*}
 &R_{i,k}= B_{{\rm c},i}\log_2\left(1+\frac{P_{i,k} \beta_{{\rm c},i}}{d_{i,k}^2(\mathbf{q}_i)}\right), \forall k\in\mathcal{K},\forall i\in\mathcal{I},
\end{align*}
where  {$B_{{\rm c},i}$ is $U_i$'s fixed} communication bandwidth and $\beta_{{\rm c},i}$ is {the} reference channel gain\footnote{The reference channel gain is the channel power gain at a reference distance of 1 m.} divided by the noise variance. 
\section{Problem Formulation}\label{section:problem formulation}
In this paper, we aim at minimizing the \ac{psl} of $\Sigma_{\rm n}(n)$ by jointly optimizing the \ac{uav} formation $\mathcal{Q}=\{\mathbf{q}_i,\forall i \in \mathcal{I}\}$ and communication transmit powers  $\mathcal{P}=\{\mathbf{P}_i, \forall i \in \mathcal{I}\}$, while satisfying communication and sensing quality-of-service constraints. To this end, we formulate the following optimization problem: 
\begin{alignat*}{2} 
&(\mathrm{P}):\;\min_{\mathcal{Q},\mathcal{P}} \hspace{3mm}  \mathrm{PSL}   & \qquad&  \\
\text{s.t.} \hspace{3mm} &  \mathrm{C1}: \; \Sigma_{\rm n}(0)\geq 1-\epsilon,               &      &  \\  & \mathrm{C2}: \; \delta_{\rm n} \leq \delta_{\rm n}^{\max},   \delta_{\rm r} \leq \delta_{\rm r}^{\max},           &      &      \\ &
\mathrm{C3}: \; z_{\mathrm{min}} \leq z_i \leq z_{\mathrm{ max}}, \forall i  \in \mathcal{I},               &      &  \\ &  \mathrm{C4}: \;  \theta_{\mathrm{min}} \leq \theta_i \leq \theta_{\mathrm{max}}, \forall i \in \mathcal{I},            &      &  \\&  \mathrm{C5}:  \;  b_{i,j} \geq d_{\mathrm{min}},\forall i,j  \in \mathcal{I}, i>j,  &      &     
 \\
 &  \mathrm{C6}: \; S_i\geq S_{\rm min}, \forall i \in \mathcal{I},        &      &     
 \\
  &  \mathrm{C7}: \; \mathrm{SNR}_{i,j} \geq \mathrm{SNR}_{\mathrm{min}}, \forall i,j \in \mathcal{I}, i\geq j,        &      &     
 \\ 
& \mathrm{C8}: \; 0 \leq P_{i,k}  \leq P_{\mathrm{max}}, \forall i \in \mathcal{I}, \forall k\in\mathcal{K},    & &\\
& \mathrm{C9}: \; R_{i,k} \geq R_{\mathrm{min}}, \forall i \in \mathcal{I}, \forall k\in\mathcal{K},       & &  \\
& \mathrm{C10}: \; E_i \leq E_{\rm max}, \forall i \in \mathcal{I}.   & &  
\end{alignat*}
Constraint $\mathrm{C1}$ enforces that the value of the \ac{psf} at the target location is close to $1$ by choosing $\epsilon\in[0,0.05]$, thereby ensuring distortion-free reconstruction of the target. Constraints $\mathrm{C2}$ impose a maximum tomographic and slant-range resolutions, denoted by $\delta_{\rm n}^{\max}$ and $\delta_{\rm r}^{\max}$, respectively. Constraint $\mathrm{C3}$ defines the permissible flight altitude range, bounded by $z_{\mathrm{min}}$ and $z_{\mathrm{max}}$. Constraint $\mathrm{C4}$ imposes limits on the allowable look angle, specified by $\theta_{\rm min}$ and $\theta_{\rm max}$.
Constraint $\mathrm{C5}$ guarantees a minimum separation distance $d_{\mathrm{min}}$ between any pair of \acp{uav}. Constraint $\mathrm{C6}$ enforces a minimum radar swath width $S_{\mathrm{min}}$. Constraint $\mathrm{C7}$ ensures that both the monostatic and bi-static \acp{snr} satisfy a minimum threshold $\mathrm{SNR}_{\mathrm{min}}$. Constraint $\mathrm{C8}$ limits the maximum communication transmit power to $P_{\mathrm{max}}$. Constraint $\mathrm{C9}$ ensures that each UAV $U_i, \forall i \in \mathcal{I}$, achieves a minimum required communication rate $R_{\mathrm{min}}$. Finally, constraint $\mathrm{C10}$ restricts the total communication energy consumption of each UAV, defined as $E_i = \delta_t \sum_{k=1}^{K} P_{i,k}$, to not exceed $E_{\mathrm{max}}, \forall i \in \mathcal{I}$.

Problem $\mathrm{(P)}$ is a challenging min-max optimization problem that is difficult to solve. In particular, the \ac{psf} expression is difficult to optimize due to its complex structure (see \eqref{eq:psf}, \eqref{eq:psf_r}, and \eqref{eq:psl}). Moreover, several constraints are non-convex, e.g.,  $\mathrm{C1}$, $\mathrm{C2}$, $\mathrm{C5}$, $\mathrm{C6}$, and $\mathrm{C9}$. Lastly, constraints $\mathrm{C2}$ are intractable, since a closed-form expression for the zeros of the \ac{psf} in \eqref{eq:resolutions} is not available, which makes conventional optimization techniques inapplicable. 
\section{Solution of the Optimization Problem}
Given the non-tractability, complexity, and non-convexity of problem $\mathrm{(P)}$, we approach it with stochastic optimization techniques. Motivated by the success of \ac{pso} methods in handling high-dimensional and complex optimization problems, we propose a modified \ac{pso} algorithm specifically tailored to solve problem $\mathrm{(P)}$. 
\subsection{Particle Swarm Optimization}\label{subsec: pso}
\ac{pso} is a population-based stochastic optimization algorithm inspired by the collective behavior of biological swarms (e.g., flocks of birds) \cite{pso}. Thus, a population of size $O \in \mathbb{N}$ explores the search space by iteratively updating a set of candidate solutions, also referred to as particles. In iteration $m \in \mathbb{N}$, each particle is represented by $\mathbf{p}^{(m)} = (p^{(m)}[1], \ldots, p^{(m)}[{D}])^{\rm T} \in \mathbb{R}^{{D}}$, where ${D}$ is the problem dimension. Each particle is updated using a velocity vector $\mathbf{v}^{(m)} = (v^{(m)}[1], \dots, v^{(m)}[{D}])^{\rm T}\in \mathbb{R}^{{D}}$ as follows:
\begin{equation}\label{eq:population_update}
\mathbf{p}^{(m)} = \mathbf{p}^{(m-1)} + \mathbf{v}^{(m-1)}, \quad m \geq 2,
\end{equation}
where initial particles $\mathbf{p}^{(1)}$ are uniformly distributed in $[\mathbf{p}^{\min},\mathbf{p}^{\max}]$, such that $\mathbf{p}^{\min}, \mathbf{p}^{\max} \in \mathbb{R}^{{D}}$ define the minimum and maximum boundary area for the \ac{pso} algorithm \cite{pso}, respectively. The initial velocities are uniformly sampled as $\mathbf{v}^{(1)}\sim\mathcal{U}(\mathbf{0},v_{\text{PSO}}^{\rm max}\mathbf{1})$, where $v_{\text{PSO}}^{\rm max}$ is the maximum particle velocity.
Particles are evaluated in each iteration using a predefined fitness function $F$, designed to meet the objective of problem $\mathrm{(P)}$. The particle velocity is updated in each iteration $m$ as follows:
\ifonecolumn
\begin{equation}\label{eq:pso_velocity_update}
\mathbf{v}^{(m)} =
w^{(m-1)} \mathbf{v}^{(m-1)}
+ c_1 \mathbf{r}_1^{(m)} \odot
\left(
\mathbf{p}_{\mathrm{best}}^{(m-1)}
- \mathbf{p}^{(m-1)}
\right)
+ c_2 \mathbf{r}_2^{(m)} \odot
\left(
\mathbf{g}_{\mathrm{best}}^{(m-1)}
- \mathbf{p}^{(m-1)}
\right),
\end{equation}
\else{\small\begin{align}\label{eq:pso_velocity_update}
\mathbf{v}^{(m)} &=
w^{(m-1)} \mathbf{v}^{(m-1)}
+ c_1 \mathbf{r}_1^{(m)} \odot
\left(
\mathbf{p}_{\mathrm{best}}^{(m-1)}
- \mathbf{p}^{(m-1)}
\right) \\ \notag
&+ c_2 \mathbf{r}_2^{(m)} \odot
\left(
\mathbf{g}_{\mathrm{best}}^{(m-1)}
- \mathbf{p}^{(m-1)}
\right),
\end{align}}\fi
where $c_1$ and $c_2$ denote the cognitive and social learning
factors, respectively. $\mathbf{r}_1^{(m)}$ and
$\mathbf{r}_2^{(m)}$ are vectors whose elements are independently
drawn from a uniform distribution over $[0,1]$, $w^{(m-1)} \in [0,1]$ is the linearly
decreasing inertia weight, $\mathbf{p}_{\mathrm{best}}^{(m-1)}$
denotes the particle's previous best-known position, and
$\mathbf{g}_{\mathrm{best}}^{(m-1)}$ represents the global
best-known position \cite{pso}.
 In this work, we adopt a reflecting-wall mechanism and set the search boundaries based on constraints $\mathrm{C3}$ and $\mathrm{C4}$ to ensure their feasibility and to accelerate convergence as described in \cite{amine5}. Readers are referred to \cite{pso} for further details about the implementation of the \ac{pso} algorithm.
\subsection{Proposed Solution}
{To solve problem $\mathrm{(P)}$, we reduce the problem dimension from $D\!=\!I(K+2)$ to $2I$ by observing that the \ac{psl} is independent of $\mathcal{P}$ and, thus, for a given \ac{uav} formation $\mathcal{Q}$, problem $\mathrm{(P)}$ becomes a feasibility problem \ac{wrt} power allocation $\mathcal{P}$: 
\begin{alignat*}{2} 
	&\mathrm{(P^\prime)}:\min_{\mathcal{P}} \hspace{3mm}  1   & \qquad&  \\
	\text{s.t.} \hspace{3mm} &\mathrm{C8,C9,C10}.&       &     
\end{alignat*}
We can reformulate constraints $\rm C8$, $\rm C9$, and $\rm C10$ as follows \cite{amine5}:
\begin{equation*}\label{eq:equivalent_communication_costraints}
    \begin{dcases}
        P_{i,k}\leq P_{\rm max}, & \forall k \in \mathcal{K}, \forall i \in \mathcal{I},\\
        P_{i,k}\geq \eta_{i,k}, & \forall k \in \mathcal{K}, \forall i \in \mathcal{I},\\
        \sum_{k=1}^{K} P_{i,k} \leq \frac{E_{\rm max}}{\delta_{\rm t}}, & \forall i \in \mathcal{I},
    \end{dcases}
\end{equation*}
where $\eta_{i,k} = \frac{d_{i,k}^2}{\beta_{{\rm c},i}}\left(2^{\frac{R_{\mathrm{min}}}{B_{{\rm c},i}}} - 1\right)$. The minimum and maximum communication transmit power of $U_i$ in time slot $k$ are denoted by $\eta_{i,k}$ and $P_{\rm max}$, respectively, and the total communication power of $U_i$ must not exceed $E_{\rm max}/\delta_{\rm t}$. We can set the transmit power consumed by $U_i$ to its minimum required level to obtain a trivial solution to the resource allocation problem as follows:
    \begin{equation}
	\mathbf{P}^*_{i} = (\eta_{i,1}, \ldots,\eta_{i,K})^{\rm T}, \quad \forall i.
	\label{eq:communication_power}
\end{equation}
Therefore, constraints $\rm C8$, $\rm C9$, and $\rm C10$ are feasible if and only if the following constraints are met:
\begin{align}\label{eq:communication_constarints}
	\begin{dcases}
		\mathrm{C11:} \ \eta_{i,k} \leq P_{\rm max}, & \forall k \in \mathcal{K}, \forall i \in \mathcal{I},\\
		\mathrm{C12:} \ \sum_{k=1}^{K} \eta_{i,k} \leq \frac{E_{\rm max}}{\delta_{\rm t}}, & \forall i \in \mathcal{I}.
	\end{dcases}
\end{align}
In other words, it suffices to check if the values $\eta_{i,k}$ satisfy constraints $\rm C11$ and $\rm C12$. If the constraints are met, $\eta_{i,k},\forall i,\forall k$, are adopted as the solution to problem $(\rm P^\prime)$. Otherwise, the difference from $P_{\rm max}$ is added to the fitness function as shown below.} Based on this formulation, the \ac{pso} search space can be reduced such that each particle encodes only the \ac{uav} positions, i.e., $ \mathcal{Q}$, while the power allocation $\mathcal{P}$ is determined separately. In other words, in iteration $m$, a particle is given by $\mathbf{p}^{(m)}=\left(\mathbf{q}^{\mathrm{T}}_{1,m}, \ldots,\mathbf{q}^{\mathrm{T}}_{I,m}\right)^{\mathrm{T}} \in\mathbb{R}^{2I}$, where $\mathbf{q}_{i,m}$  is the candidate across-track position of $U_i$. Therefore, for a candidate \ac{uav} formation selected by $\mathbf{p}^{(m)}$, the power allocation $\mathcal{P}^*=\{\mathbf{P}^*_{i}, \forall i\}$ is computed based on (\ref{eq:communication_power}) and used to evaluate $\mathbf{p}^{(m)}$ using the following non-parameterized fitness function \cite{genocopII}:
\begin{equation}\label{eq:fitness_function_s1} F(\mathbf{p}^{(m)} | \mathcal{P}^*) = \begin{dcases} \mathrm{PSL}(\mathbf{p}^{(m)}), & \text{if } \mathbf{p}^{(m)} \in \mathcal{F},\vspace{4mm}\\ \mathrm{PSL}_{\rm max} + \sum_{l \in \mathcal{L}} g_l (\mathbf{p}^{(m)} | \mathcal{P}^*), & \text{otherwise}, \end{dcases} \end{equation}
where $\mathcal{F}$ denotes the feasible set, $\mathrm{PSL}_{\rm max}$ is the worst \ac{psl} across the population, and $g_l, l\in\mathcal{L},\mathcal{L}=\left\{1, 2, 4, 5, 6, 7, 11,12\right\}$, quantify constraint violations as follows:
\ifonecolumn
\begin{align*}
	g_1 &= [1 - \epsilon-\Sigma_{\rm n}(0)]^+,\\
    g_2 &= [\delta_{\rm r} - \delta_{\rm r}^{\rm max}]^+ + [\delta_{\rm n} - \delta_{\rm n}^{\rm max}]^+,\\
	g_4 &= \sum_{i=1}^I \left([\theta_{\rm min} - \theta_i]^+ + [\theta_i - \theta_{\rm max}]^+\right),\\
	g_5 &= \sum_{i>j} [d_{\rm min} - b_{i,j}]^+,\\
	g_6 &= \sum_{i=1}^I [S_{\rm min} - S_i]^+,\\
	g_7 &= \sum_{i=1}^I \sum_{j \le i} [\mathrm{SNR}_{\rm min} - \mathrm{SNR}_{i,j}]^+,\\
    g_{11} &= \sum_{i=1}^I \sum_{k=1}^K [\eta_{i,k} - P_{\rm max}]^+,\\
	g_{12} &= \sum_{i=1}^I \sum_{k=1}^K \left[\eta_{i,k} - \frac{E_{\rm max}}{\delta_{\rm t}}\right]^+.
\end{align*}
\else
{\begin{align*}
	&g_1 = [1 - \epsilon-\Sigma_{\rm n}(0)]^+,
    g_2 = [\delta_{\rm r} - \delta_{\rm r}^{\rm max}]^+ + [\delta_{\rm n} - \delta_{\rm n}^{\rm max}]^+,\\
	&g_4 = \sum_{i=1}^I \left([\theta_{\rm min} - \theta_i]^+ + [\theta_i - \theta_{\rm max}]^+\right),
	g_5 = \sum_{i>j} [d_{\rm min} - b_{i,j}]^+,\\
	&g_6 = \sum_{i=1}^I [S_{\rm min} - S_i]^+,
	g_7 = \sum_{i=1}^I \sum_{j \le i} [\mathrm{SNR}_{\rm min} - \mathrm{SNR}_{i,j}]^+,\\
    &g_{11} = \sum_{i=1}^I \sum_{k=1}^K [\eta_{i,k} - P_{\rm max}]^+,
	g_{12} = \sum_{i=1}^I \sum_{k=1}^K \left[\eta_{i,k} - \frac{E_{\rm max}}{\delta_{\rm t}}\right]^+.
\end{align*}}\fi
The resulting proposed \ac{pso}-based algorithm is summarized in \textbf{Algorithm} \ref{alg:pso}. 
\ifonecolumn
\begin{algorithm}[t]
	\caption{Proposed PSO-based Solution}\label{alg:pso}
	\begin{algorithmic}[1]
		\Statex \textbf{Output:} Solution to Problem $\mathrm{(P)}$ 
		\Statex\textbf{Initialization:}
		\State\hspace{\algorithmicindent}Initialize iteration $m \gets 1$, random \ac{pso} population of $O$ particles, and random \ac{pso} velocities.  
		\State\hspace{\algorithmicindent}Set the best particle based on the fitness function $F$ defined in (\ref{eq:fitness_function_s1}).
		\While{$m \leq M$}
		\State Update PSO population and velocities using (\ref{eq:population_update}) and (\ref{eq:pso_velocity_update}), respectively.
        \State Compute $\mathcal{P}^*$ for each particle $\mathbf{p}^{(m)} \in \mathbb{R}^{2I}$ using \eqref{eq:communication_power} and (\ref{eq:communication_constarints}).
		\State Evaluate the fitness of each particle $\mathbf{p}^{(m)}$ given  $\mathcal{P}^*$ using \eqref{eq:fitness_function_s1}.
		\State Update and record the best particles $\mathbf{p}_{\mathrm{best}}^{m}$ and their respective fitness and the worst fitness value \Statex\hspace{\algorithmicindent}$\mathrm{PSL}_{\mathrm{max}}$.
		\State Update iteration number $m \gets m + 1$.
		\EndWhile
		\State \textbf{return} The best \ac{uav} formation $\mathbf{p}_{\mathrm{best}}^{M}$ and its associated communication powers $\mathcal{P}^*$.
	\end{algorithmic}
\end{algorithm}
\else\begin{algorithm}[t]
	\caption{Proposed PSO-based Solution}\label{alg:pso}
	\begin{algorithmic}[1]
		\Statex \textbf{Output:} Solution to Problem $\mathrm{(P)}$ 
		\Statex\textbf{Initialization:}
		\State\hspace{\algorithmicindent}Initialize iteration $m \gets 1$, random \ac{pso} population of
        \Statex\hspace{\algorithmicindent}$O$ particles, and random \ac{pso} velocities.  
		\State\hspace{\algorithmicindent}Set the best particle based on the fitness function 
        \Statex\hspace{\algorithmicindent}defined in (\ref{eq:fitness_function_s1}).
		\While{$m \leq M$}
		\State Update PSO population and velocities 
        \Statex\hspace{\algorithmicindent}using  (\ref{eq:population_update})
        and (\ref{eq:pso_velocity_update}), respectively.
        \State Compute $\mathcal{P}^*$ for each particle $\mathbf{p}^{(m)}$ using \eqref{eq:communication_power} and (\ref{eq:communication_constarints}).
		\State Evaluate the fitness of each particle $\mathbf{p}^{(m)}$ given  $\mathcal{P}^*$
        \Statex\hspace{\algorithmicindent}using \eqref{eq:fitness_function_s1}.
		\State Update and record the best particles $\mathbf{p}_{\mathrm{best}}^{m}$
        and their \Statex\hspace{\algorithmicindent}respective fitness and the worst fitness value $\mathrm{PSL}_{\mathrm{max}}$.
		\State Update  iteration number $m \gets m + 1$.
		\EndWhile
		\State \textbf{return} The best \ac{uav} formation $\mathbf{p}_{\mathrm{best}}^{M}$ and its associated \Statex\hspace{\algorithmicindent}\hspace{\algorithmicindent}communication powers $\mathcal{P}^*$.
	\end{algorithmic}
\end{algorithm}
\fi
\section{Simulation Results}\label{sec:simulation_results}
This section presents simulation results for the proposed \ac{pso}-based algorithm outlined in \textbf{Algorithm}~\ref{alg:pso}. Unless otherwise specified, the parameters used are as listed in Table~\ref{tab:my-table}. The algorithm was run for a maximum of $M = 500$ iterations with $O=500$ particles. The learning factors are set to $c_1=2$ and $c_2=2.5$, and the maximum \ac{pso} velocity is constrained to $v_{\mathrm{PSO}}^{\mathrm{max}} = 1$.
\subsection{Benchmark Schemes}
To evaluate performance, we compare the proposed solution with the conventional \ac{pso} algorithm as well as two state-of-the-art evolutionary algorithms as benchmark schemes to solve problem $(\rm P)$: 
\subsubsection{\ac{cga} \cite{cga}} 
This benchmark scheme employs \ac{cga} to solve problem~$\mathrm{(P)}$, incorporating a natural selection mechanism with a selection rate of $0.5$, Gaussian mutation with a mutation rate of $0.1$, and blend crossover (\(\alpha_b\)-BLX) with \(\alpha_b = 0.3\). The algorithm is executed for a maximum of $300$ generations. For brevity, the implementation details of the \ac{cga} are omitted, and interested readers are referred to~\cite{cga} for a comprehensive description.
\begin{table}[]
	\centering
	\caption{System parameters \cite{victor2,snr_equation,coherence1}.}
	\label{tab:my-table}
	\begin{adjustbox}{max width=\columnwidth}
		\begin{tabular}{|c|c?c|c?c|c|}
			\hline
			Parameter           & Value 					& Parameter & Value 	& Parameter &Value  \\ \hline
			$I$ & $6$ & $h_{\rm min}$ & $5$ m&$\sigma_0$  &$-10$ dB	\\ \hline 
			$K$  &$200$ &$P_{\mathrm{max}}$ & $10$ dB &$P_{\rm t}$  & $10$ dB\\ \hline
			$\delta_{\rm t}$&  $1$ s& $E_{\rm max}$ & $570$ J	& $G_{\rm t}$  &$5$ dBi    \\ \hline
			$z_{\mathrm{min}}$& $1$ m &$B_{\mathrm{c},i}, \forall i$ & $1$ GHz  & $G_{\rm r}$   &$5$ dBi  \\ \hline
			$z_{\mathrm{max}}$&  $100$ m  &$\beta_{\mathrm{c},i}, \forall i$ & $20$ dB &$\lambda$   &$12$ cm \\ \hline
			$\theta_{\mathrm{min}}$ & $37.24^\circ$ 
			&$g_{\rm x}$  &$-85$ m  &$L$ &$6$ dB   \\ \hline
			$\theta_{\rm max}$ & $48.7^\circ$ & $g_{\rm y}$   &$400$	m
			&$T_{\rm sys}$  &$400$ k \\ \hline
			$v_{\rm y}$ & $4.3$ m/s&$g_{\rm z}$    & $25$ m  &  $F$  & $5$ dB \\ \hline
			$x_{\rm t}$ & $20$ m &$R_{\rm min}$ & $6$ Mb/s &$B_{\mathrm{noise}}$ & $3$ GHz \\ \hline 
			$d_{\rm min}$    & $2$ m&$\delta_{\rm n}^{\rm max}$ & $1$ m   & $\theta_{3\mathrm{dB}}$ & $40^\circ$  \\ \hline
			$S_{\rm min}$ & $55$ m&$\delta_{\rm r}^{\rm max}$ & $20$ cm & $\mathrm{SNR}_{\mathrm{min}}$ & $-10$ dBm  	  \\ \hline
		\end{tabular}
	\end{adjustbox} \vspace{-4mm}
\end{table}
\subsubsection{\ac{genocop} \cite{genocopII}}
\Ac{genocop} is an extension to the Genocop I algorithm \cite{genocopI}, designed for constrained optimization problems. We set the initial temperature to $10^5$. The uniform, non-uniform, and boundary rates are set to $0.2$, $0.7$, and $0.7$, respectively. The population size is $100$, and the maximum number of generations for Genocop I and \ac{genocop} is $300$ and $100$, respectively. Detailed descriptions of Genocop I and \ac{genocop} can be found in \cite{genocopI} and \cite{genocopII}, respectively. 
\subsubsection{Particle Swarm Optimization \cite{pso}}
In this scheme, we implement the conventional \ac{pso} algorithm using the same parameters used in the proposed solution but without reducing the problem dimension, i.e., a particle $\mathbf{p}^{(m)}\in\mathbb{R}^{(K+2)I}$ encodes both the \ac{uav} formation and communication powers as follows:
\ifonecolumn\begin{equation*}
\mathbf{p}^{(m)}=\left(\underbrace{p^{(m)}[1],\cdots,p^{(m)}[2I]}_{\mathcal{Q}},\underbrace{p^{(m)}[2I+1],\cdots,p^{(m)}[(K+2)I]}_{\mathcal{P}}\right)^{\rm T}.
\end{equation*}
\else{\small\begin{equation*}
\mathbf{p}^{(m)}=(\underbrace{p^{(m)}[1],\cdots,p^{(m)}[2I]}_{\mathcal{Q}},\underbrace{p^{(m)}[2I+1],\cdots,p^{(m)}[D]}_{\mathcal{P}})^{\rm T}.
\end{equation*}}\fi
\subsection{Numerical Results}
\begin{figure}[t!]
	\centering
	\ifonecolumn
	\includegraphics[width=4in]{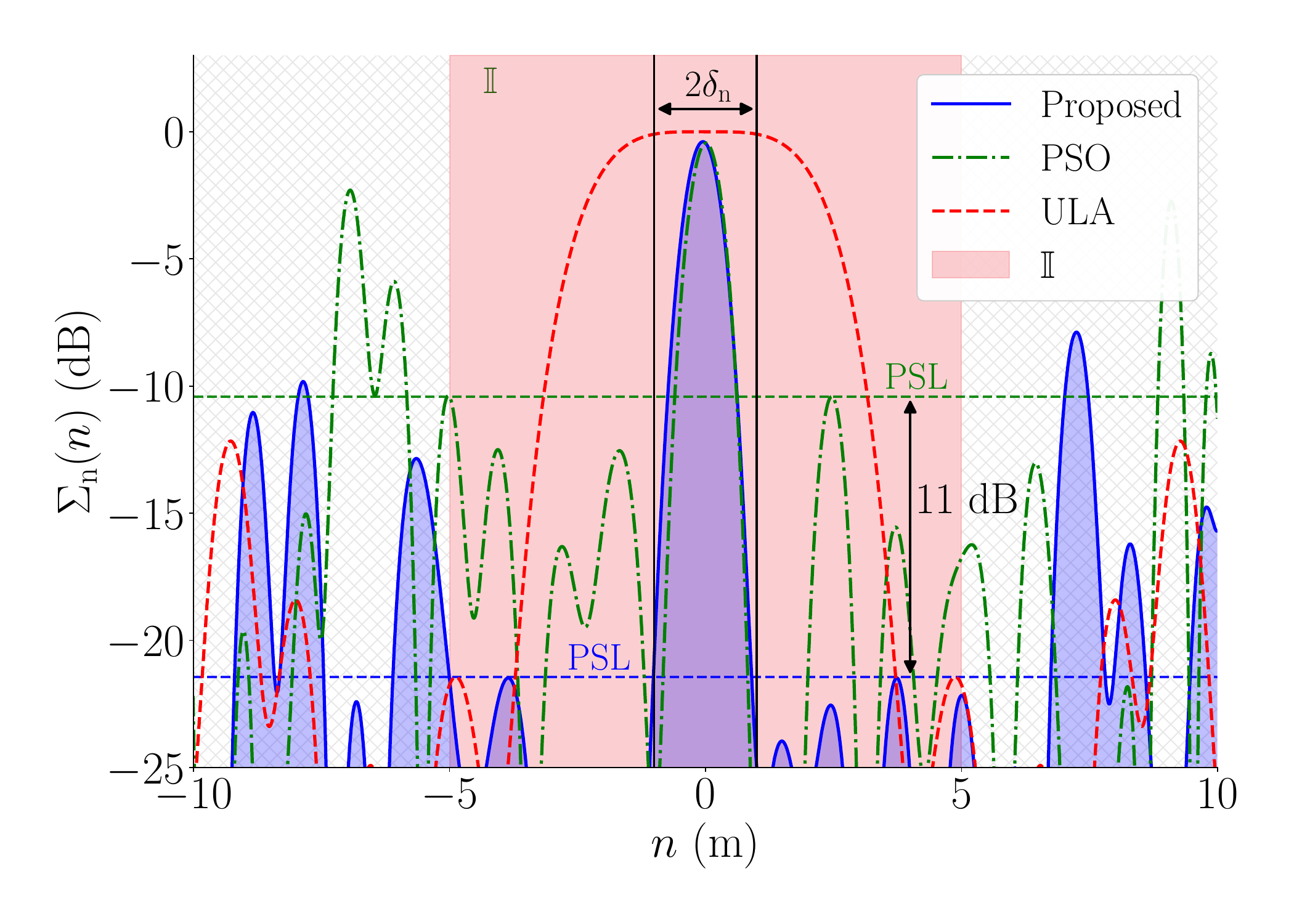}
	\else
	\includegraphics[width=1\columnwidth]{figures/psf_I=6_hmin=5.pdf}
	\fi
	\caption{Comparison of the optimized \ac{psf}, $\Sigma_{\rm n}(n)$, obtained using the proposed scheme, the \ac{pso} benchmark, and a conventional \ac{ula} for $h_{\rm max}=5$ m and $\delta_{\rm n}=1$ m.}
	\label{fig:psf}
\end{figure}

Fig.~\ref{fig:psf} shows the optimized elevation \ac{psf}, $\Sigma_{\rm n}(n)$, obtained using the proposed scheme and the conventional \ac{pso} algorithm. In addition, the \ac{psf} of a \ac{ula}, which is commonly used for performing \ac{tomosar}, with a uniform \ac{uav} spacing of $12.6$ m, is included for comparison. It can be observed that, by effectively reducing the of the search space, the proposed scheme achieves a performance gain of $11$ dB compared to the \ac{pso} benchmark scheme. 
Furthermore, although the \ac{ula} achieves a \ac{psl} comparable to that of the proposed scheme, this performance is attained at the cost of a significantly wider mainlobe, resulting in reduced height resolution.

\begin{figure}[t!]
	\centering
	\ifonecolumn
	\includegraphics[width=4in]{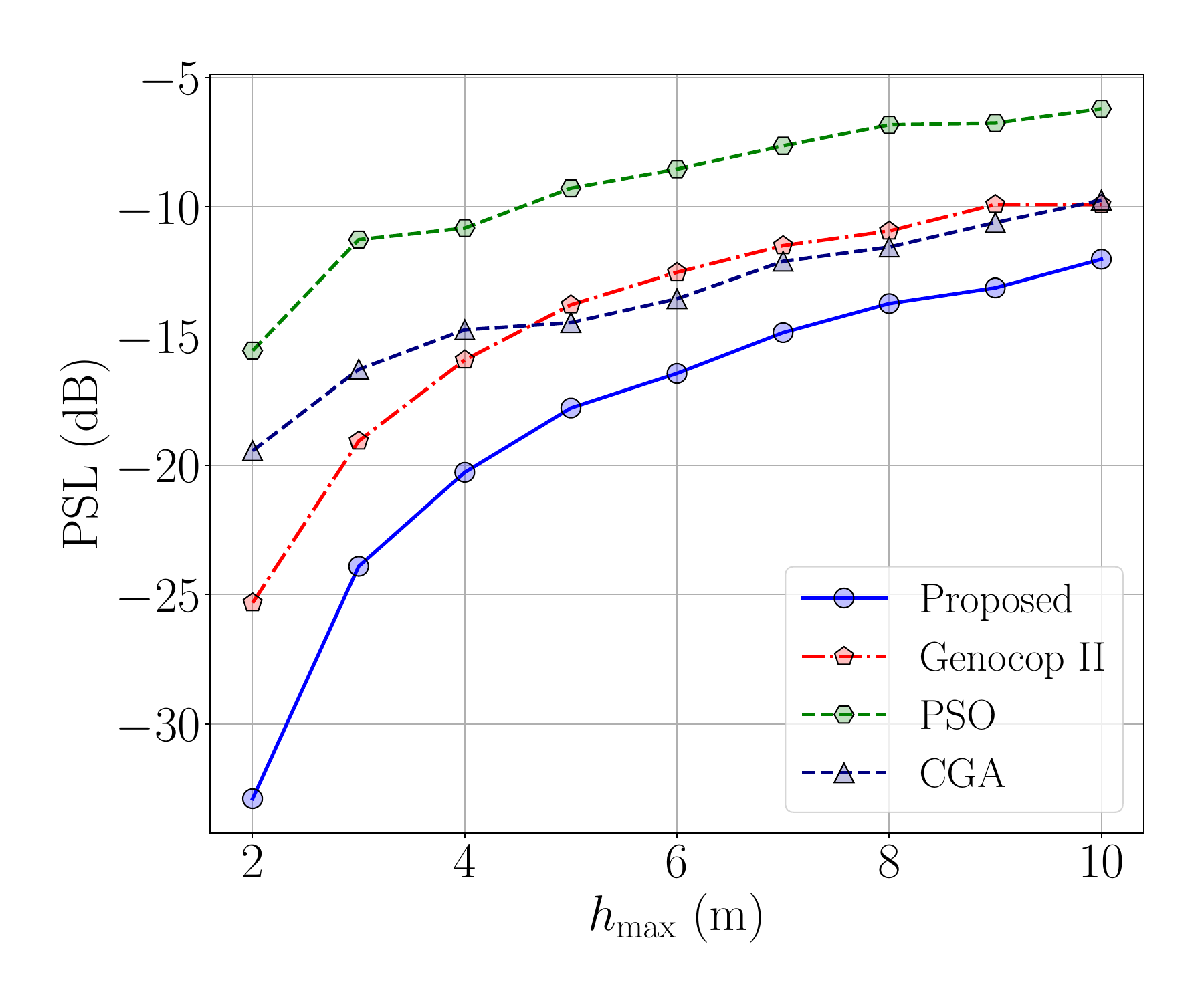}
	\else
	\includegraphics[width=0.9\columnwidth]{figures/objective_vs_hmin.pdf}
	\fi
	\caption{ \ac{psl} achieved using the proposed scheme and the benchmark methods versus different maximum height $h_{\rm max}$.}
	\label{fig:objective_vs_hmin}
\end{figure}

Fig.~\ref{fig:objective_vs_hmin} shows the \ac{psl} of $\Sigma_{\rm n}(n)$ achieved by the proposed approach and the benchmark schemes for different maximum heights, $h_{\rm max}$. It can be observed that the \ac{psl} increases as $h_{\rm max}$ increases. This indicates a fundamental trade-off, whereby achieving a larger $h_{\rm max}$ results in more pronounced sidelobes.
Among the benchmark methods, both \ac{genocop} and \ac{cga} outperform the conventional \ac{pso}, but neither matches the performance of the proposed scheme.
This performance gain can be attributed to the reduction of variable coupling through the restriction of the search space, which enables more effective sidelobe suppression.

\begin{figure}[t!]
	\centering
	\ifonecolumn
	\includegraphics[width=4in]{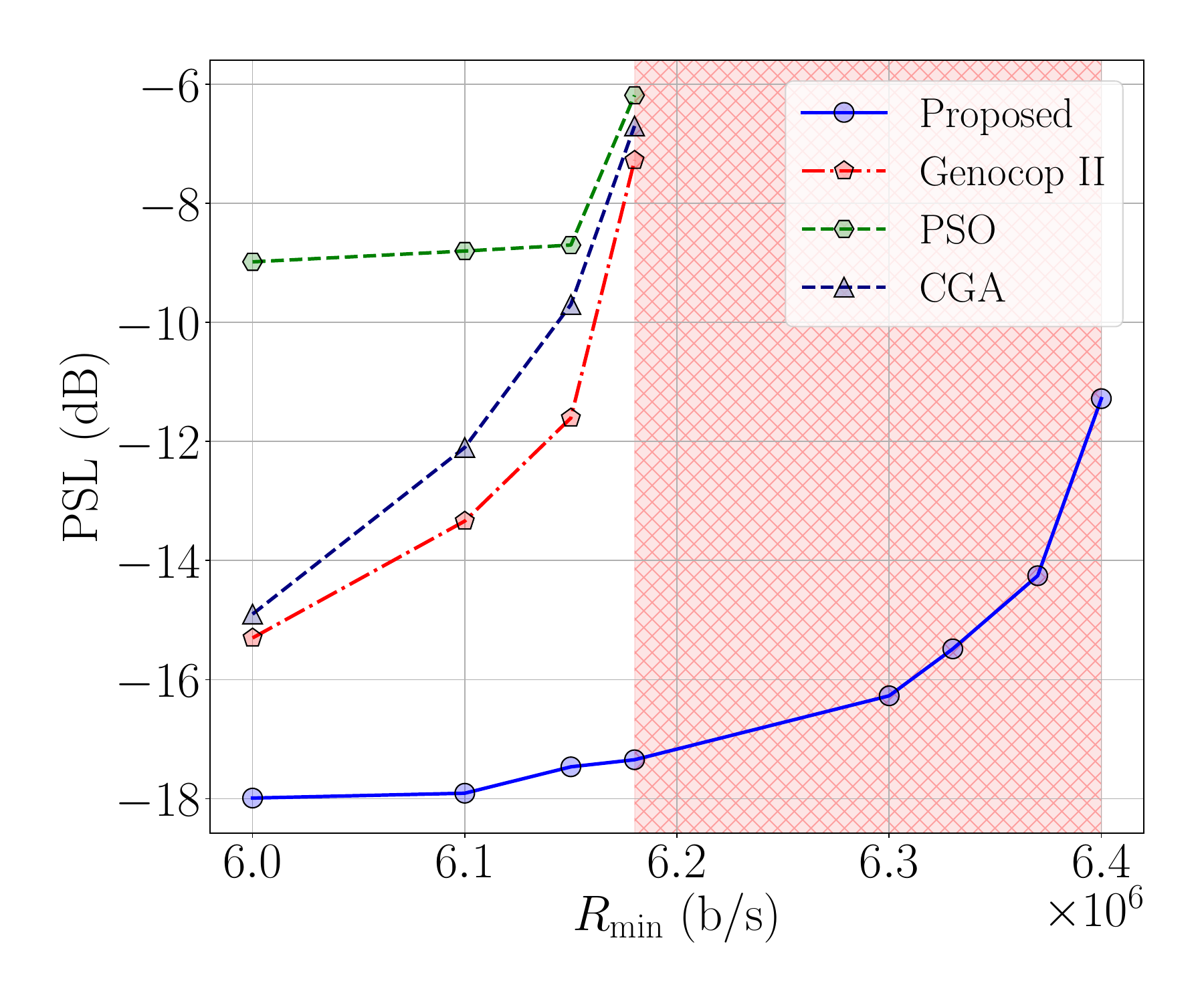}
	\else
	\includegraphics[width=0.9\columnwidth]{figures/objective_vs_Rmin.pdf}
	\fi
	\caption{\ac{psl} achieved using the proposed scheme and the benchmark methods versus different minimum required data rate $R_{\rm min}$.}
	\label{fig:objective_vs_Rmin}
\end{figure}

Fig.~\ref{fig:objective_vs_Rmin} shows the \ac{psl} as a function of the minimum required data rate, $R_{\rm min}$. It can be observed that more stringent data rate requirements lead to higher sidelobes. This can be attributed to the fact that higher data rate requirements impose stricter constraints on the distances between $U_i, \forall i$ and the \ac{gs}, which in turn activates constraint $\rm C9$.
Moreover, the infeasibility of all benchmark schemes for $R_{\rm min} > 6.18$ Mb/s underscores the critical role of efficient communication power allocation in enabling real-time data offloading. Finally, Fig.~\ref{fig:objective_vs_Rmin} demonstrates that the proposed scheme consistently outperforms all benchmark approaches.

\section{conclusion}
In this work, we investigated a \ac{uav}-borne \ac{mimo} \ac{tomosar} system to produce a \ac{3d} image of an area by acquiring $I^2$ independent observations from different angles. We formulated an optimization problem aimed at minimizing the \ac{psl} of the \ac{psf}, accounting for the relevant sensing and communication quality-of-service constraints, thereby enhancing the quality of the reconstructed image. We proposed a modified \ac{pso}-based approach that jointly optimizes the \ac{uav} formation as well as the communication power allocation, while fulfilling sensing and communication constraints. Simulation results showed that the proposed scheme significantly improves the \ac{psl} suppression compared to several benchmark schemes.

\bibliographystyle{IEEEtran}
\bibliography{biblio}

@book{book1,
	title={Introduction to Radar Systems},
	author={Skolnik, Merrill Ivan and others},
	volume={3},
	year={1980},
	publisher={McGraw-hill New York}
}

@article{coherence1,
	title = {Coherence evaluation of {TanDEM-X} interferometric data},
	journal = {ISPRS J. Photogramm. Remote Sens.},
	volume = {73},
	pages = {21-29},
	year = {2012},
	note = {},
	issn = {0924-2716},
	doi = {https://doi.org/10.1016/j.isprsjprs.2012.06.006},
	url = {},
	author = {Michele Martone and others}
}

@INPROCEEDINGS{amine3,
	author={Lahmeri, Mohamed-Amine and others},
	booktitle={Proc. IEEE Int. Conf. Commun.}, 
	title={{UAV} Formation Optimization for Communication-Assisted {InSAR} Sensing}, 
	year={2024},
	volume={},
	number={},
	pages={3913-3918},
	doi={10.1109/ICC51166.2024.10622647}}

@ARTICLE{stso,
	author={Krieger, Gerhard},
	journal={IEEE Trans. Geosci. Remote Sens.}, 
	title={{MIMO-SAR}: Opportunities and Pitfalls}, 
	year={2014},
	volume={52},
	number={5},
	pages={2628-2645},
	doi={10.1109/TGRS.2013.2263934}}

@Article{seker,
	author = {Seker, Ilgin and Lavalle, Marco},
	title = {Tomographic Performance of Multi-Static Radar Formations: Theory and Simulations},
	journal = {Remote Sensing},
	volume = {13},
	year = {2021},
	number = {4},
	ARTICLE-NUMBER = {737},
	DOI = {10.3390/rs13040737}
}

@book{cga,
	title={Practical Genetic Algorithms},
	author={Haupt, Randy L and Haupt, Sue Ellen},
	year={2004},
	publisher={John Wiley \& Sons}
}

@article{genocopI,
	title={{GENOCOP}: a genetic algorithm for numerical optimization problems with linear constraints},
	author={Michalewicz, Zbigniew and Janikow, Cezary Z.},
	journal={Commun. ACM},
	volume={39},
	number={12},
	pages={175--185},
	year={1996},
	publisher={ACM}
}

@inproceedings{genocopII,
	author    = {Z. Michalewicz and N. Attia},
	title     = {Evolutionary Optimization of Constrained Problems},
	booktitle = {Proc. 3rd Annu. Conf. Evol. Programming},
	publisher = {World Scientific Publishing},
	year      = {1994},
	pages     = {98--108}
}

@INPROCEEDINGS{pso,
	author={Kennedy, J. and Eberhart, R.},
	booktitle={Proc. Int. Conf. Neural Netw.}, 
	title={Particle swarm optimization}, 
	year={1995},
	volume={4},
	number={},
	pages={1942-1948}}

@ARTICLE{victor2,
	author={Mustieles-Perez, Victor and others},
	journal={IEEE Geosci. Remote Sens. Lett.}, 
	title={New Insights Into Wideband Synthetic Aperture Radar Interferometry}, 
	year={2024},
	volume={21},
	number={},
	pages={1-5}}

@ARTICLE{snr_equation,
	author={Krieger, Gerhard and others},
	journal={IEEE Trans. Geosci. Remote Sens.}, 
	title={{TanDEM-X}: A Satellite Formation for High-Resolution {SAR} Interferometry}, 
	year={2007},
	volume={45},
	number={11},
	pages={3317-3341},
	doi={10.1109/TGRS.2007.900693}}

@misc{amine5,
	title={Sensing Accuracy Optimization for Multi-{UAV SAR} Interferometry with Data Offloading}, 
	author={Mohamed-Amine Lahmeri and others},
	year={2025},
	eprint={2507.11284},
	archivePrefix={arXiv},
	primaryClass={eess.SP},
	url={https://arxiv.org/abs/2507.11284}, 
}

@ARTICLE{tutorial,
	author={Moreira, Alberto and others},
	journal={IEEE Geosci. Remote Sens. Mag.}, 
	title={A tutorial on synthetic aperture radar}, 
	year={2013},
	volume={1},
	number={1},
	pages={6-43},
	doi={10.1109/MGRS.2013.2248301}}

@ARTICLE{reigber,
	author={Reigber, A. and Moreira, A.},
	journal={IEEE Trans. Geosci. Remote Sens.}, 
	title={First demonstration of airborne {SAR} tomography using multibaseline {L}-band data}, 
	year={2000},
	volume={38},
	number={5},
	pages={2142-2152},
	doi={10.1109/36.868873}}

@INPROCEEDINGS{kreiger,
	author={Krieger, Gerhard and Rommel, Tobias and Moreira, Alberto},
	booktitle={Proc. 11th Europ. Conf. on Synthetic Aperture Radar}, 
	title={{MIMO-SAR} Tomography}, 
	year={2016},
	volume={},
	number={},
	pages={1-6},
	doi={}}

@INPROCEEDINGS{amine6,
  author={Lahmeri, Mohamed-Amine and others},
  booktitle={Proc. IEEE Int. Conf. Commun.}, 
  title={Sensing Accuracy Optimization for Communication-Assisted Dual-Baseline {UAV-InSAR}}, 
  year={2025},
  volume={},
  number={},
  pages={6573-6578},
  doi={10.1109/ICC52391.2025.11161780}}

@ARTICLE{fornaro,
	author={Fornaro, G. and Serafino, F. and Soldovieri, F.},
	journal={IEEE Trans. Geosci. Remote Sens.}, 
	title={Three-dimensional focusing with multipass {SAR} data}, 
	year={2003},
	volume={41},
	number={3},
	pages={507-517},
	doi={10.1109/TGRS.2003.809934}}

@ARTICLE{zhu,
	author={Zhu, Xiao Xiang and Bamler, Richard},
	journal={IEEE Trans. Geosci. Remote Sens.}, 
	title={Very High Resolution Spaceborne {SAR} Tomography in Urban Environment}, 
	year={2010},
	volume={48},
	number={12},
	pages={4296-4308},
	doi={10.1109/TGRS.2010.2050487}}

@ARTICLE{rambour,
  author={Rambour, Clement and others},
  journal={IEEE Geosci. Remote Sens. Mag.}, 
  title={From Interferometric to Tomographic {SAR}: A Review of Synthetic Aperture Radar Tomography-Processing Techniques for Scatterer Unmixing in Urban Areas}, 
  year={2020},
  volume={8},
  number={2},
  pages={6-29},
  doi={10.1109/MGRS.2019.2957215}}

@book{fornaro2,
  title={Multi-Dimensional Imaging with Synthetic Aperture Radar: Theory and Applications},
  author={Fornaro, Gianfranco and others},
  year={2024},
  publisher={Academic Press},
}

\end{document}